# CONSIDERATIONS REGARDING THE MODELLING OF WIND ENERGY CONVERSION SYSTEMS

**I. Szeidert, O. Prostean, I. Filip, N. Budisan**
Department of Automation and Industrial Informatics
Faculty of Automation and Computer Science
Bd. V. Parvan, No.2, 1900 Timisoara, ROMANIA
Tel: +40 256 403237  Fax: +40 256 403214  E-mail: siosif@aut.utt.ro

*Abstract:* *The above paper presents some considerations regarding the modelling of wind energy conversion systems (WECS). There are presented practical problems of grid integration of wind turbines, the usage of general system models, respectively of RMS (root mean squares) models.*
*The are presented models of some WECS components and related, such as: a probabilistic 2D model for instantaneous wind velocities, aerodynamic model of wind turbine, rotating inertia mode, asynchronous machine model, grid model.*
*This paper only presents models used in different WECS, models that can be easily simulated with adequate simulation tools such as Matlab – Simulink.*

## 1. Introduction

### 1. 1. Practical problems in grid integration of wind turbines

Practical problems in grid integration of wind turbines define special cases in power system analysis and design. These problems are very often characterised by
- Power systems which are composed of equipment and components from different fields of engineering, such as electrical, mechanical, thermodynamic, automatic control and other related
fields. This means for a comprehensive analysis approach that all relevant components have to be considered with a similar level of detail and system models can not be restricted to electrical components primarily.
- Time dependent multidimensional system environment. This means the simulation of the system environment should be part of the analysis approach. This subject becomes difficult if random and additionally correlated environment parameter disturb the power system. [6]

### 1.2. The Engineering point of view

Phenomena in power systems are related to different **types of parameter**, **time domain and location** as it is well known from empirical observations. A very rough systematisation distinguishes these phenomena
with respect to the following dimensions:
- the type of the relevant power system parameter (electromagnetic, mechanic, thermodynamic etc.)
- the relevance of the location in continua (magnetic-, electric-field etc.)
- the behaviour of the parameter in time (instantaneous-, mean-, root-mean-square (RMS) values).

Models and methods developed in power system engineering have been oriented at the different character of the targeted area. Within the MATLAB product-family the Power System Engineering Toolset, Part RMSDynamic Simulation is devoted to the RMS-Dynamic of power systems and the Power System Blockset targets the Transient Dynamic primarily.

An additional aspect of phenomena in power systems is related to the **3-phase character** of the electrical equipment because of the dominating nature of three phase power systems.

Models for transient-dynamic simulations are based on differential equations for instantaneous values and distinguish between the different phases of the electrical equipment. [2] [4]

Models for RMS-Dynamic simulations neglect the transients in the grid related to leakage flux linkages and electric fields but not those transients related to the rotor-flux linkages of rotating electrical machine. This approach leads to models for the

electrical components of the power system with the following main characteristics:
- the grid including the stator of the rotating machine will be represented by algebraic complex phasor equations (models in frequency domain which considers only one phase)
- the rotor of the rotating machine will be represented by ordinary differential equations (models in time domain, distinguishing different phases).

The different modelling approaches express a theoretical difference behind the Power System Engineering Toolset and the Power System Blockset.

### 1.3. The mathematical point of view

The general concept which guidelines the approach applied in this paper in a formal sense is the concept of time-scale-modelling. If continua phenomena will be neglected, the system-model will have the following principle structure in time domain:
- sets of ordinary and mostly non-linear differential equations which can be expressed in the state space form model the power system
- deterministic functions depend on time as well as multidimensional and multivariate random processes in time model the system environment.

Models of this type can be linearised for model analysis purposes.

The most important consequence dealing with models of this type is the necessity to solve linear or nonlinear sets of equations during the integration-process of the differential equations.

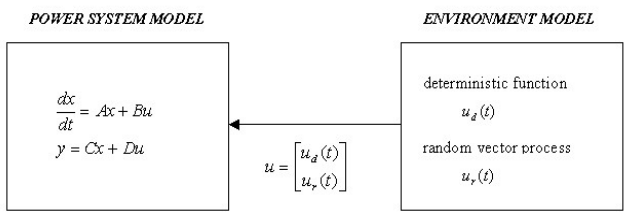

Figure 1. General model structure.

The environment of the power systems will be modelled using time series for electrical loads (real and reactive power) as well as for real environmental parameter like the wind speed. In case of deterministic functions the time series will directly be calculated from those functions. (Figure 1).

In case of random time series (e.g. the wind speeds affecting wind turbines) the problem is much more difficult because of the necessity to match specific random properties of those parameter sufficiently. There are three general requirements in this respect:
- each time series shell fulfil a specified probability density functions (PDF) as well as a specified power spectral density functions (PSD)
- between two time series a specified cross spectral density function (CSD), expressed via the coherence function and the angle of the complex transfer function shell be fulfilled
- the time series should be steady state (with respect to their random properties) from the start of the simulation to avoid a not desired length of the simulation time.

The final structure of power system models in the RMS-time domain will have the following general formal structure (Figure2):

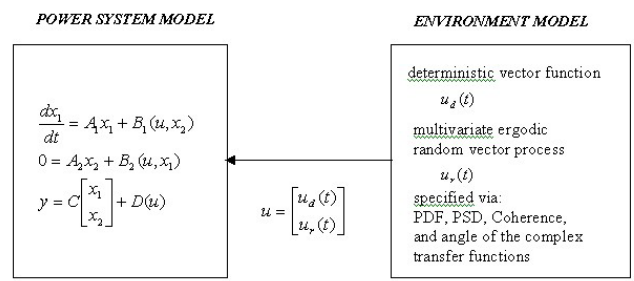

Figure 2. General structure of RMS-Models

## 2. Theoretical description of specific models used

### 2.1 Probabilistic 2D model for instantaneous wind velocities

In Figure 3 and Figure 4 is presented a probabilistic 2D model for instantaneous wind velocities.

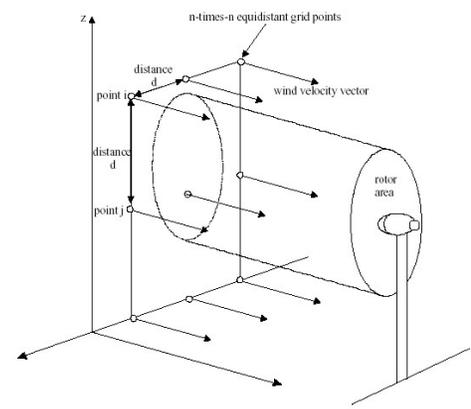

Figure 3. Probabilistic 2D model for instantaneous wind velocities

An ergodic random process model can be specified by:
· Probability Density Function (PDF) for wind velocity in each grid point
· Power Spectral Density function (PSD) for wind velocity in each grid point
· Spectral Coherence function (COH) for wind velocities in two grid points
· Spectral angle of transfer function (angle_TF) for wind velocities in two grid points

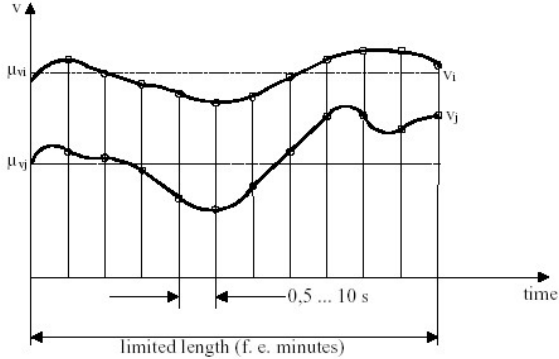

Figure 4. Principle of two sampled random wind velocities in grid points i and j with limited length (minutes)

As an example nearby is presented a PDF-model:
· Gaussian distribution (uniform for all grid points)

$$f(v) = \frac{1}{\sqrt{2\pi\sigma_v^2}} \exp\left[-\frac{(v-\mu_v)^2}{2\sigma_v^2}\right] \quad (1)$$

Height-Profile for mean velocity

$$\mu_v = v_H \left(\frac{Z}{Z_N}\right)^\alpha \quad (2)$$

Direct turbulence intensity specification

$$\sigma_v = \mu_v R \quad (3)$$

Panowsky-turbulence model

$$\sigma_v = \frac{\mu_v}{\ln\left(\frac{z}{z_0}\right)} \quad (4)$$

The afferent variables and parameters are:
$f$ [s/m] probability density
$v$ [m/s] wind velocity
$\sigma_v$ [m/s] standard deviation of the wind velocity
$\mu_v$ [m/s] mean of the wind velocity
$R$ [-] turbulence intensity
$z$ [m] height above ground
$z_0$ [m] roughness length of the terrain
$Z_N$ [m] nacelle height
$v_m$ [m/s] wind velocity in nacelle height

**2.2 Aerodynamic model of wind turbine – rotor - Coordinate systems**

The wind park coordinates (WP) are shown in Figure 5.

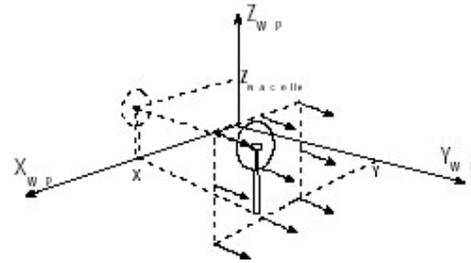

Figure 5. Wind Park Coordinates (WP)

Vector of wind velocity $\vec{v}_W = \begin{vmatrix} 0 \\ v_W \\ 0 \end{vmatrix}$ (5)

Position $\quad Position, WP = \begin{vmatrix} X \\ Y \\ Z \end{vmatrix}$ (6)

The wind turbine coordinates are ilustrated in Figure 6.

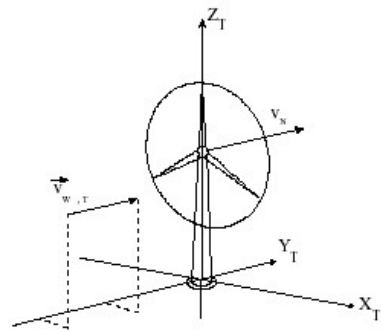

Figure 6. Wind Turbine Coordinates.

Vector of wind velocity

$$\vec{v}_{W,T} = \vec{v}_W = \begin{vmatrix} 0 \\ v_W \\ 0 \end{vmatrix} \quad (7)$$

Position

$$Position, T = Position, WP - \begin{vmatrix} X \\ Y \\ 0 \end{vmatrix} = \begin{vmatrix} 0 \\ 0 \\ Z \end{vmatrix} \quad (8)$$

Turbine Disc Coordinates (TD) are ilustrated in Figure 7.

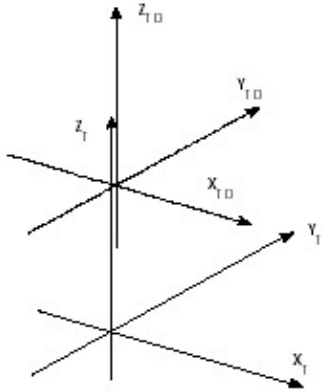

Figure 7. Turbine Disc Coordinates

Vector of wind velocity $\quad \vec{v}_{W,TD} = \vec{v}_{W,T} = \begin{vmatrix} 0 \\ v_W \\ 0 \end{vmatrix} \quad (9)$

Position $\quad Position, TD = Position, T - \begin{vmatrix} 0 \\ 0 \\ Z_{Nacelle} \end{vmatrix} \quad (10)$

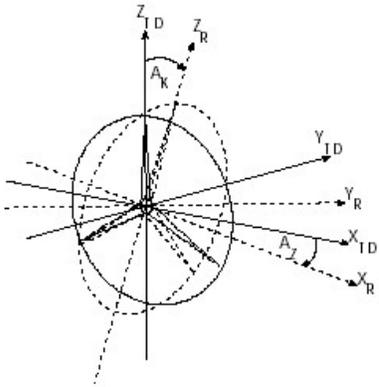

Figure 8. Rotor coordinates.

The rotor coordinates are ilustrated in figure 8.
Vector of wind velocity

$$\vec{v}_{W,R} = T_{EL} T_{AZ} \vec{v}_{W,T} \quad (11)$$

Position $\quad Position, R = T_{EL} T_{AZ} Position, TD \quad (12)$
with

$$T_{EL} = \begin{vmatrix} 1 & 0 & 0 \\ 0 & \cos A_K & -\sin A_K \\ 0 & \sin A_K & \cos A_K \end{vmatrix} \quad (13)$$

$$T_{AZ} = \begin{vmatrix} \cos A_Z & -\sin A_Z & 0 \\ \sin A_Z & \cos A_Z & 0 \\ 0 & 0 & 1 \end{vmatrix} \quad (14)$$

### 2.3. Model of rotating inertia

In the following figure is presented the scheme of stiff rotating inertia:

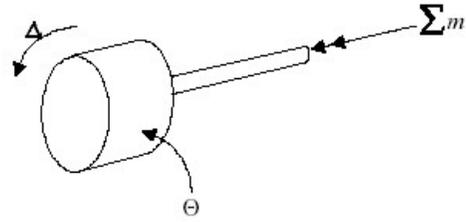

Figure 9. Scheme of stiff rotating inertia
Dynamic model:

$$\Theta \frac{d^2 \Delta}{dt^2} = -k_f \frac{d\Delta}{dt} + \sum m \quad (15)$$

where: $\Theta$ [kg m$^2$] inertia; $\Delta$ [rad] angle; $k_f$ [Nm/(rad/s)] friction factor; m[Nm] shaft torque

### 2.4. Shaft/Gearbox model

In figure 10 is presented the model principle scheme of a gearbox from a WECS.

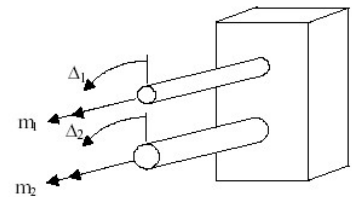

Figure 10. Shaft/Gearbox model principle scheme.
Torque model:

$$\begin{bmatrix} m_1 \\ m_2 \end{bmatrix} = \begin{bmatrix} -1 & \frac{1}{n} \\ n & \frac{-1}{n^2} \end{bmatrix} \left( c \begin{bmatrix} \Delta_1 \\ \Delta_2 \end{bmatrix} + d \frac{d}{dt} \begin{bmatrix} \Delta_1 \\ \Delta_2 \end{bmatrix} \right) \quad (16)$$

where: m [Nm] shaft torque; Δ [rad] shaft angle; c [Nm/rad] stiffness constant; d [Nm/(rad/s)] damping constant; n [-] transformation ratio

## 2.5. Asynchronous machine model

General winding scheme related to each pole pair is presented in following figure.

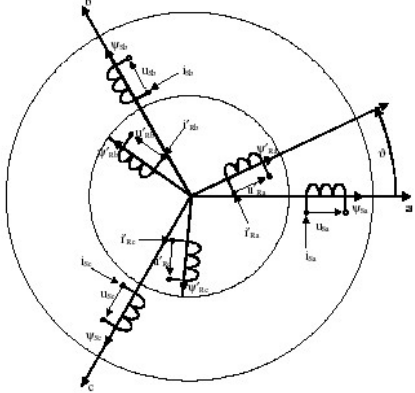

Figure 11. Single cage / slipring machine – winding scheme.

Voltage equations:
$$\begin{bmatrix} u_S \\ u'_R \end{bmatrix} = \begin{bmatrix} R_{SS} & 0 \\ 0 & R'_{RR} \end{bmatrix} \begin{bmatrix} i_S \\ i'_R \end{bmatrix} + \frac{d}{dt} \begin{bmatrix} \psi_S \\ \psi'_R \end{bmatrix} \quad (17)$$

Flux linkage equation:
$$\begin{bmatrix} \psi_S \\ \psi'_R \end{bmatrix} = \begin{bmatrix} L_S & L_{SR} \\ L_{RS} & L'_R \end{bmatrix} \begin{bmatrix} i_S \\ i'_R \end{bmatrix} \quad (18)$$

Electromagnetic shaft torque:
$$m = \frac{p}{\sqrt{3}} (T_S \psi_S)^T i_S \quad (19)$$

The variables and parameters of the asynchronous machine are:

m [Nm] electromagnetic shaft toque; p [-] number of pole pairs; $u_S$ [V] instantaneous stator voltage; $u'_R$ instantaneous rotor voltage (stator related); T [-] transpose operator; $\psi_S$ [Vs] instantaneous stator flux linkage; $\psi'_R$ [Vs] instantaneous rotor flux linkage (stator related);

## 2.6. Grid model

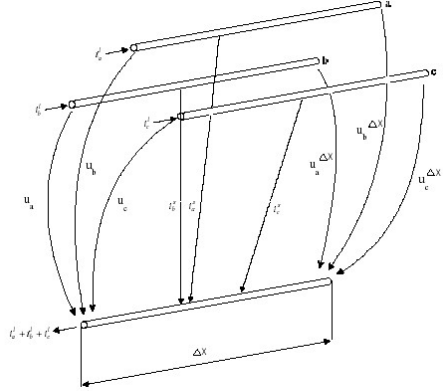

Figure 12. General scheme of a line segment.

Variables and parameters: a, b, c [-] phase; $\Delta X$ [m] length of wire segment

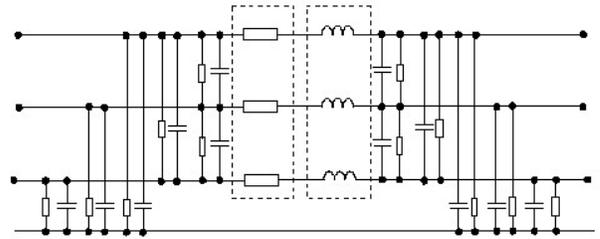

Figure 13. Equivalent circuit representation using concentrated parameters (6$^{th}$ order model).

Voltage / current equations:

$$L' \frac{d}{dt} \begin{bmatrix} i^l_a \\ i^l_b \\ i^l_c \end{bmatrix} + R' \begin{bmatrix} i^l_a \\ i^l_b \\ i^l_c \end{bmatrix} = \frac{1}{\Delta X} \begin{bmatrix} u_a - u_a^{\Delta x} \\ u_b - u_b^{\Delta x} \\ u_c - u_c^{\Delta x} \end{bmatrix} \quad (20)$$

$$\begin{bmatrix} C'_E + 2C'_L & -C'_L & -C'_L \\ -C'_L & C'_E + 2C'_L & -C'_L \\ -C'_L & -C'_L & C'_E + 2C'_L \end{bmatrix} \frac{d}{dt} \begin{bmatrix} u_a \\ u_b \\ u_c \end{bmatrix}$$
$$+ \begin{bmatrix} G'_E + 2G'_L & -G'_L & -G'_L \\ -G'_L & G'_E + 2G'_L & -G'_L \\ -G'_L & -G'_L & G'_E + 2G'_L \end{bmatrix} \begin{bmatrix} u_a \\ u_b \\ u_c \end{bmatrix} = \frac{1}{\Delta X} \begin{bmatrix} i^s_a \\ i^s_b \\ i^s_c \end{bmatrix}$$
(21)

$$L' = T^{-1} \begin{bmatrix} L'_0 & & \\ & L'_1 & \\ & & L'_1 \end{bmatrix} T \quad R' = T^{-1} \begin{bmatrix} R'_0 & & \\ & R'_1 & \\ & & R'_1 \end{bmatrix} T$$
(22-23)

$$T = \frac{1}{3}\begin{bmatrix} 1 & 1 & 1 \\ 1 & \exp j\frac{2}{3}\pi & \exp j\frac{4}{3}\pi \\ 1 & \exp j\frac{4}{3}\pi & \exp j\frac{2}{3}\pi \end{bmatrix} \quad (24)$$

Variables and parameters:
$R'_0\,[\Omega/m]$ zero sequence resistance per meter; $R'_1\,[\Omega/m]$ positive sequence resistance per meter; $\Delta X\,[m]$ length of wire segment; $G'_E\,[S/m]$ earth conductance per meter; $G'_L\,[S/m]$ line conductance per meter; $C'_E\,[F/m]$ earth capacitance per meter; $C'_L\,[F/m]$ line capacitance per meter; $L'_0\,[H/m]$ zero sequence inductance per meter; $L'_1\,[H/m]$ positive sequence inductance per meter;

**2.7. Example of an WECS (Wind Energy Conversion System)**

**Wind turbine Vestas V27 with blade element rotor model and cp-lambda rotor model**

Principal characteristics:
· Asynchronous generator
· Capacitor bank
· Pitch controller of the wind turbine rotor for power limited
· Voltage and frequency modifications are possible
· Stochastic wind field

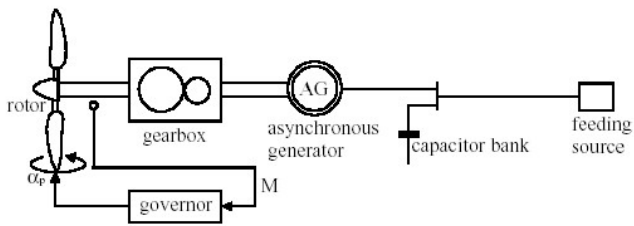

Figure 14. Principle representation – Vestas V27 – wind turbine.

### 3. Conclusions

The models presented in the prevoius paragraphes are the models of a WECS componenets: wind turbine, gearbox, electrical machine. In order to validate the models through simulation, the WECS has been completed with the wind model and respectively a grid model. Using the described models it is possible to study different functioning regimes, to study the transient regimes,etc.

The presented models can be easily implemented in Maltlab-Simulink in order to simulate the behavoiur and evaluate different WECS.

There is also presented an implemented WECS – the Vestas V-27 wind turbine.